\documentclass[
    ,final            
  ]
  {aipproc}

\layoutstyle{8x11single}

\begin{document}

\title{Hydrodynamic Flow in PbPb Collisions Observed via Azimuthal Angle Correlations of Charged Hadrons}


\author{Eric Appelt for the CMS Collaboration}{
address={Vanderbilt University,
6301 Stevenson Center, 
VU Station B No.351807,
Nashville, TN 37235, United States}
}

\begin{abstract}
Azimuthal angle correlations of charged hadrons were measured in $\sqrt s_{NN}$ = 2.76 TeV PbPb collisions 
by the CMS experiment. The distributions exhibit anisotropies that are correlated with the event-by-event 
orientation of the reaction plane. Several methods were employed to extract the strength of the signal: 
the event-plane, cumulant and Lee-Yang Zeros methods. These methods have different sensitivity to 
correlations that are not caused by the collective motion in the system (non-flow correlations due to jets, 
resonance decays, and quantum correlations). The second Fourier coefficient of the charged hadron 
azimuthal distributions was measured as a function of transverse momentum, pseudorapidity and 
centrality in a broad kinematic range: $0.3 < p_T < 12.0$ GeV/c, $|\eta| < 2.4$, as a function of 
collision centrality. In addition, the third through sixth Fourier components were measured at 
midrapidity using selected methods.
\end{abstract}

\maketitle


\section{Introduction}

In non-central heavy ion collisions, the initial spatial asymmetry of the collision zone leads to anisotropies in the final state 
charged hadron emission. This collective behavior is known as ``flow'' and is the result of the hydrodynamic expansion of the matter produced in the collision zone and is driven by uneven pressure gradients. The strength of this flow is measured through a Fourier expansion of the charged hadron azimuthal distributions with respect to the reaction plane, defined by the impact parameter vector and collision direction. The strength of the flow can be related to the transport properties of the strongly interacting medium produced in the collision.

The second Fourier coefficienct ($v_2$) has been studied extensively, while higher harmonics are more challenging 
due to smaller signals. Although the odd-order harmonics, $v_3$ and $v_5$, have been expected to vanish 
due to the symmetry of the collision system assuming a smooth spacial eccentricity profile, 
the geometric fluctuations of the positions of nucleons in the colliding nuclei may produce
event-by-event fluctuations in the initial eccentricity 
resulting in odd-order harmonics in the final state~\cite{PhysRevC.81.054905}.

Extracting the transport coefficients of the strongly coupled quark-gluon plasma from measured
flow is difficult as the initial state can only  be described by theoretical modeling. 
Different models of the initial state of the collision lead to different predictions for the strength of the
flow given the same shear and bulk viscosities of the medium~\cite{Heinz2011}. Additionally, the various methods applied
in the experimental determination of the azimuthal anisotropies have varying sensitivities to ``non-flow'' azimuthal 
correlations, such as resonance decays and jets. 


\section{Experimental Methods}

First results on elliptic and higher order flow of charged hadrons measured in PbPb collisions at $\sqrt s_{NN}$ = 2.76 TeV 
are reported. The analysis was performed using 2.31 million events selected by a minimum-bias trigger requiring a coincidence
between signals in the two forward hadronic calorimeters or the beam scintillator counters with 
an estimated trigger efficiency of ($97\pm 3$)\% of the total inelastic cross section. Further offline 
event selection was applied to remove beam-halo and beam-gas 
events, and to limit variations in tracking efficiency at high pseudorapidity. The $v_n$ coefficients were measured using the CMS
tracker comprised of silicon pixel and strip detectors covering a pseudorapidity range of $|\eta| < 2.4$. Low transverse momentum 
particles (0.3 <$p_T < 1.8$ GeV/c) were reconstructed using only the pixel detectors, while higher transverse momentum particles
($p_T > 1.5$ GeV/c) were reconstructed using the full tracker. 
A detailed description of the CMS detector~\cite{JINST} is found elsewhere.

The elliptic flow ($v_2$) analysis~\cite{CMS-PAS-HIN-10-002} was performed with the event plane 
method~\cite{PhysRevC.58.1671}, 2 and 4-particle cumulant method~\cite{PhysRevC.64.054901,Cumulant2}, 
and the Lee-Yang zeros (LYZ) method~\cite{Bhalerao2003373, 0954-3899-30-8-092}. 
Higher order flow ($v_3$ through $v_6$) analysis~\cite{CMS-PAS-HIN-11-005} was performed 
with select methods. The odd harmonics, $v_3$ and $v_5$, were measured using the 2-particle cumulant method. 
Additionally, the fourth
harmonic, $v_4$ was measured using the LYZ  method and 3 and 5-particle cumulant methods, and the sixth 
harmonic, $v_6$ was measured using the LYZ method.

For the event plane method, elliptic flow was measured with respect to event planes determined from tracks in the 
backward ($-2<\eta<-1$) and forward ($1<\eta<2$) pseudorapidity regions. This provides a gap of $ 1< \Delta\eta < 3.4$
between the tracks used in the elliptic flow measurement and those used in the event plane determination, which is expected
to reduce the effect of non-flow correlations on the measurement. 


The cumulant method measures flow by a cumulant expansion of multi-particle azimuthal correlations, 
and the flow harmonics are obtained from the cumulant terms of the expansion. 
The cumulant of the $k$-particle correlation, called the $k^{th}$ order cumulant, 
removes the contribution of all lower order correlations and so is expected to remove the effect non-flow 
correlations between less than $k$ particles.
To calculate the cumulant expansion, a generating function of the multiparticle correlations in a 
complex plane was used.
The LYZ method uses the large-order behavior of the cumulant expansion, rather than explicitly computing cumulants
at a given order.
The flow signal is obtained from the zeros in the complex plane of a generating function of azimuthal correlations.

In both the cumulant and LYZ methods, the feasibility of the measurements is dependent on the magnitude of the resolution
parameter $\chi \equiv v_n \sqrt{M}$, where $v_n$ is the harmonic used for integrated flow, or average flow over the measured
transverse momentum range, and $M$ is the multiplicity of charged hadrons. Typically, $v_n$ is too small in central collisions, and 
$M$ is too small in peripheral collisions, making the higher-order cumulant and LYZ measurements challenging under these
conditions.

The full details of the implementation of these methods for this analysis are described in references \cite{CMS-PAS-HIN-10-002}
and \cite{CMS-PAS-HIN-11-005}. 

\begin{figure}
  \includegraphics[height=.37\textheight]{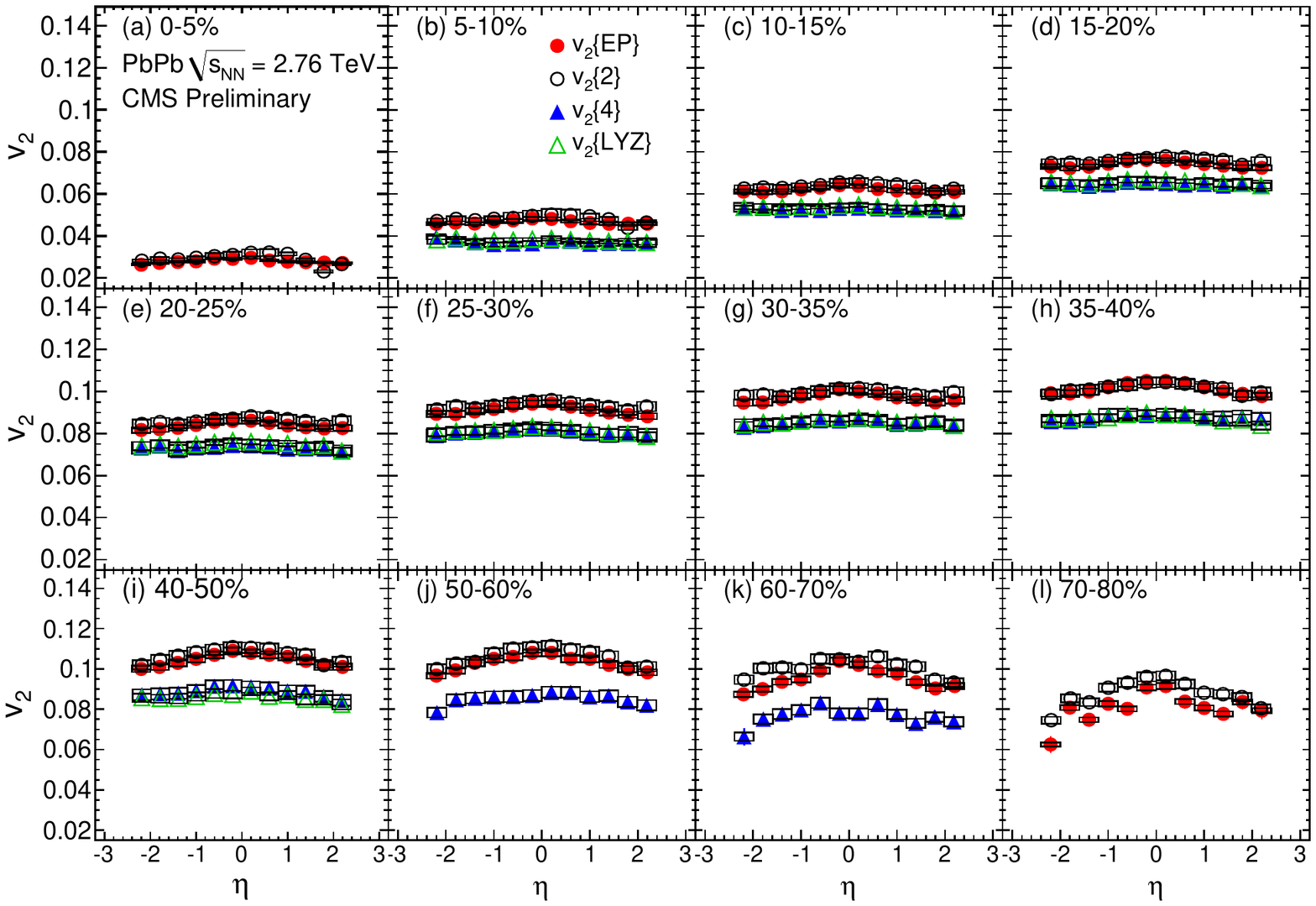}
  \caption{Comparison of the four methods for determining $v_2$ as a function of pseudorapidity $\eta$ for the 12 centrality
  classes. The $v_2$ signal has been integrated over a transverse momentum range of $0.3 < p_T < 3.0$ GeV/c. The 
  error bars show the statistical uncertainties, and the boxes represent the systematic uncertainties.\label{v2eta}}
\end{figure}

\begin{figure}
  \includegraphics[width=.57\textwidth]{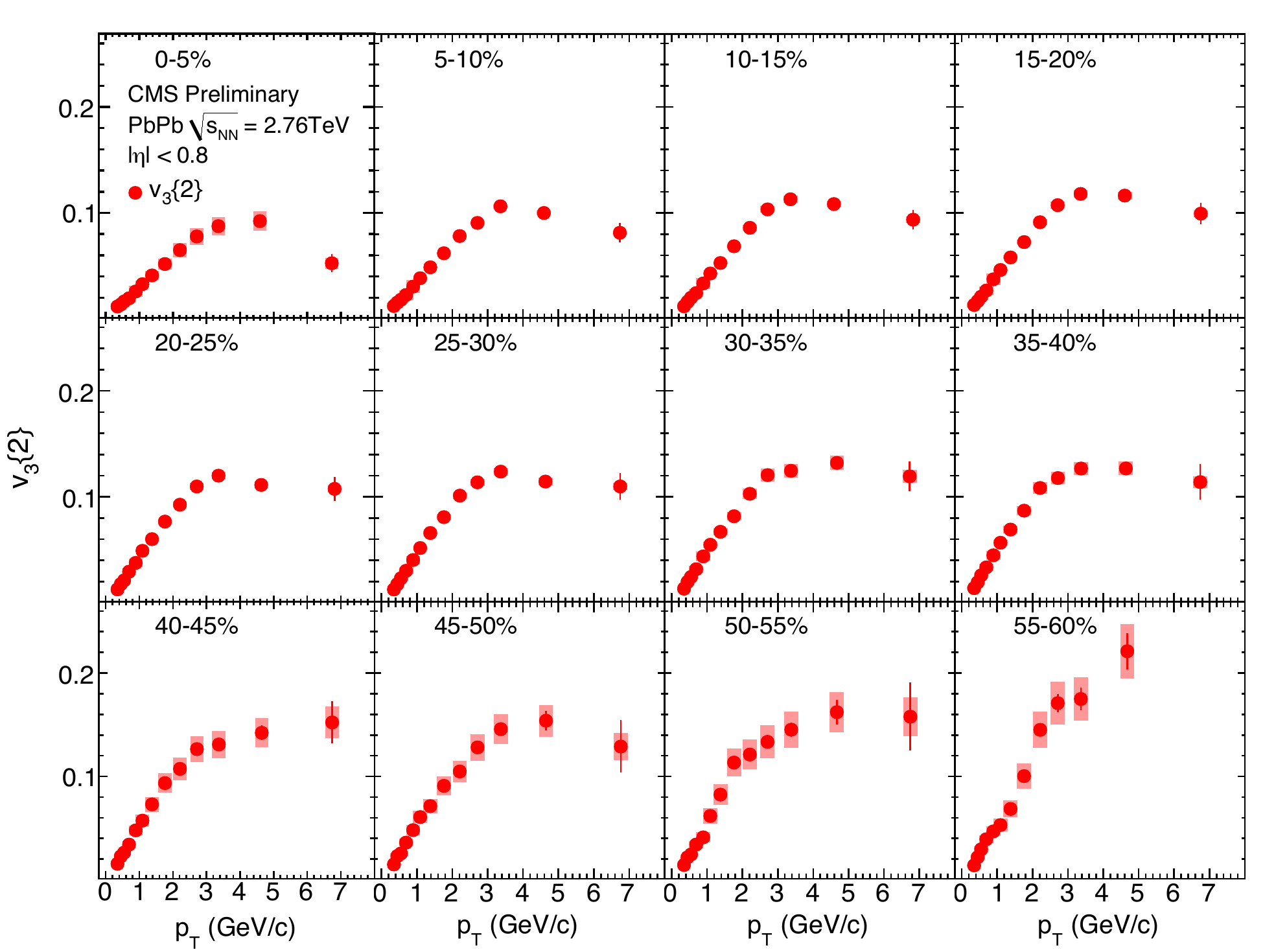}
  \includegraphics[width=.37\textwidth]{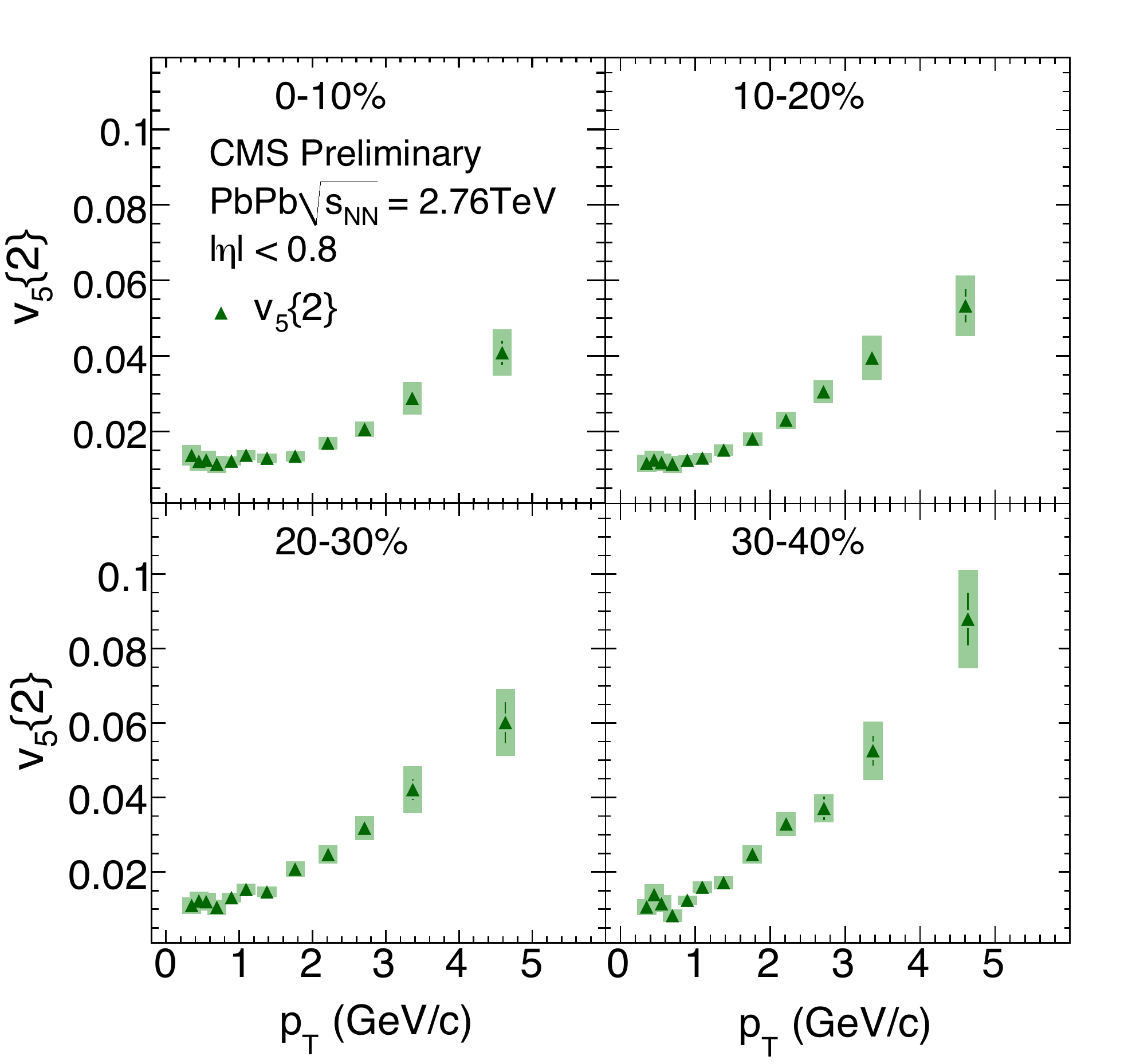}
  \caption{Odd harmonics: $v_3\{2\}$ (left) and $v_5\{2\}$ (right) as a function of $p_T$ at mid-rapidity ( $|\eta| < 0.8$)
  for the centrality classes indicated in the figures. The error bars represent the statistical uncertainties, and the colored 
  bands represent the systematic uncertainties.\label{vodd}}
\end{figure}

\section{Results and Discussion}

The main results of the analysis include: 
1) $v_n(p_T)$ at mid-rapidity $|\eta| < 0.8$ for up to 12 centrality classes in the range 0-80\%, 
2) integrated $v_2$ as a function of pseudorapidity ( $v_2(\eta)$ ) for 12 centrality classes,
and 3) integrated $v_n$ at mid-rapidity $|\eta| < 0.8$. 
Figures of the complete results and 
discussion are given in references \cite{CMS-PAS-HIN-10-002} and 
\cite{CMS-PAS-HIN-11-005}. Here, a few of the main results are shown and briefly described.

Figure~\ref{vodd} shows the odd harmonic coefficients measured as a function of $p_T$ in several centrality 
classes. One notable feature of these results is the weak centrality dependence in both the shape and magnitude
of the signals, which is expected if these harmonics are induced by fluctuations in the initial conditions that also
have a weak centrality dependence~\cite{PhysRevC.82.034913}. The $p_T$ dependence 
of the even harmonics, which are not shown here, 
exhibit behavior compatible with hydrodynamic flow.
The $v_2(\eta)$ results are presented in figure~\ref{v2eta} and show only a weak $\eta$-dependence, except in 
the most peripheral events, which are more affected by non-flow correlations. The pseudorapidity dependence may 
provide constraints on the description of the system evolution in the longitudinal direction.

These measurements taken together may aid in the systematic validation of different approaches to the modeling
of heavy ion collisions and lead to a reliable determination of some of the transport properties of the hot QCD matter
produced in the collisions.


\begin{theacknowledgments}
This research is supported in part by the Department of Energy Office of Science Graduate Fellowship Program (DOE SCGF), made possible in part by  the American Recovery and Reinvestment Act of 2009, administered by ORISE-ORAU under contract no. DE-AC05-06OR23100.

Copyright CERN for the benefit of the CMS Collaboration.
\end{theacknowledgments}



\bibliographystyle{aipproc}   

\bibliography{2K5_Appelt}

\IfFileExists{\jobname.bbl}{}
 {\typeout{}
  \typeout{******************************************}
  \typeout{** Please run "bibtex \jobname" to optain}
  \typeout{** the bibliography and then re-run LaTeX}
  \typeout{** twice to fix the references!}
  \typeout{******************************************}
  \typeout{}
 }

\end{document}